# Design, Construction and Testing of the Digital Hadron Calorimeter (DHCAL) Electronics

C. Adams[a], A. Bambaugh[d], B. Bilki[a,e], J. Butler[c], F. Corriveau[f], T. Cundiff[a], G. Drake[a], K. Francis[a,*], V. Guarino[a], B. Haberichter[a], E. Hazen[c], J. Hoff[d], S. Holm[d], A. Kreps[a], P. DeLurgio[a], L. Dal Monte[d], N. Mucia[g], E. Norbeck[e,†], D. Northacker[e], Y. Onel[e], B. Pollack[g], J. Repond[a], J. Schlereth[a], J. R. Smith[a,h,+], D. Trojand[f,^], D. Underwood[a], M. Velasco[g], J. Walendziak[a], K. Wood[a], S. Wu[c], L. Xia[a], Q. Zhang[a,b,#], A. Zhao[a]

[a] *Argonne National Laboratory, 9700 S. Cass Avenue, Argonne, IL 60439, U.S.A.*
[b] *Institute of High Energy Physics, Chinese Academy of Sciences, P.O. Box 918, Beijing, 100049, PRC*
[c] *Boston University, 590 Commonwealth Avenue, Boston, MA 02215, U.S.A.*
[d] *Fermilab, P.O. Box 500, Batavia, IL 60510-0500, U.S.A.*
[e] *University of Iowa, Iowa City, IA 52242-1479, U.S.A.*
[f] *McGill University, 3600 University Street, Montreal, QC H3A 2T8, Canada*
[g] *Northwestern University, 2145 Sheridan Road, Evanston, IL 60208-3112, U.S.A.*
[h] *University of Texas at Arlington, P.O. Box 19059, Arlington, TX 76019, U.S.A.*
*\* Now at Northern Illinois University*
*+ Now at University of Maryland*
*^ Now at University of Windsor*
*# Now at Xi'an Jiaotong University*
*†Deceased*

**Abstract.** A novel hadron calorimeter is being developed for future lepton colliding beam detectors. The calorimeter is optimized for the application of Particle Flow Algorithms (PFAs) to the measurement of hadronic jets and features a very finely segmented readout with $1 \times 1$ cm$^2$ cells. The active media of the calorimeter are Resistive Plate Chambers (RPCs) with a digital, i.e. one-bit, readout. To first order the energy of incident particles in this calorimeter is reconstructed as being proportional to the number of pads with a signal over a given threshold. A large-scale prototype calorimeter with approximately 500,000 readout channels has been built and underwent extensive testing in the Fermilab and CERN test beams. This paper reports on the design, construction, and commissioning of the electronic readout system of this prototype calorimeter. The system is based on the DCAL front-end chip and a VME-based back-end.

# INTRODUCTION

In contrast to typical calorimeters of past and present High Energy Physics Experiments with their tower structure, imaging calorimeters feature comparatively large numbers of readout channels. In these calorimeters, the cells with an area of a fraction of a 1 cm$^2$ to about 10 cm$^2$ are read out individually. The resulting fine granularity offers several distinct advantages:

- Showers can be measured individually, a prerequisite for the application of Particle Flow Algorithms (PFAs) to the measurement of hadronic jets [1];

- Electromagnetic subshowers can be identified within hadronic showers, offering the possibility to apply software compensation techniques;

- The measurements in the last layers of the calorimeter can be utilized to apply leakage correction, which in turn improve the resolution;

- $\gamma$'s, $\pi^0$'s, $e^\pm$'s, and neutral hadrons can readily be identified; and

- The direction of photons can be reconstructed.

In this context the CALICE collaboration [2] developed a Digital Hadron Calorimeter (DHCAL) based on Resistive Plate Chambers (RPCs) [3]. A large scale prototype DHCAL was built in 2008 – 2010 and was subsequently tested in the Fermilab and CERN test beams [4-6]. This paper describes the design, construction and commissioning of the electronic readout system. Details of the mechanical design, construction, and assembly of the DHCAL can be found in [7].

# DESCRIPTION OF THE DHCAL

The DHCAL prototype constitutes the first large-scale hadron calorimeter with digital readout and embedded front-end electronics. It also utilized, for the first time, a pad-readout together with RPCs. The design of the DHCAL was based on preliminary work done with a small-scale prototype, see references [8 – 13] for further details. The DHCAL was exposed to cosmic rays and to particle beams both at Fermilab and at CERN.

The DHCAL was tested in various configurations: a) at Fermilab with a 38-layer steel absorber structure followed by a steel tail catcher, b) also at Fermilab in a 50-layer structure without additional absorber plates (the skins of the detector cassettes and the RPCs themselves providing the only absorber material), and c) at CERN with a 39-layer tungsten absorber structure followed by a steel tail catcher. Figure 1 shows a photograph of the set-up at the CERN Proton-synchrotron test beam.

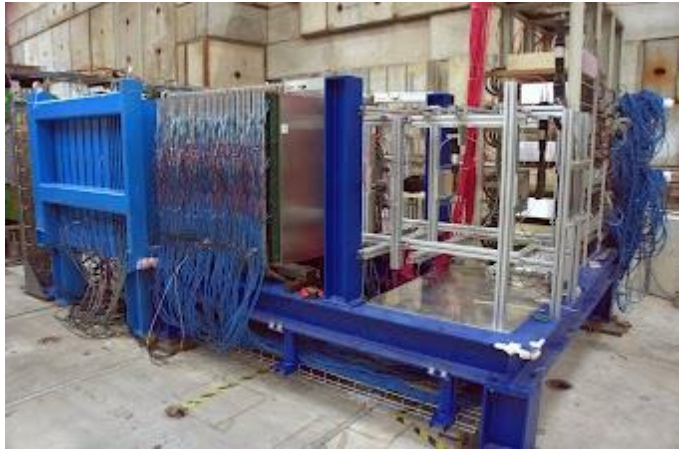

**Figure 1.** Photograph of the DHCAL setup at CERN showing the main stack with tungsten plates followed by the tail catcher with steel absorber plates.

Each layer was read out by 9216 1 × 1 cm$^2$ pads located on the bottom side of the Readout boards. To first order the charges generated by single avalanches in RPCs are determined by the ionization in the gas gap that is the furthest away from the anode and thus subject to the largest gas multiplication. This leads to a wide range of signal charges [8] which are if at all only weakly correlated to the energy loss in the gas gap. Therefore, a precise measurement of the induced charge is seen as only of limited interest. As a consequence, the DHCAL readout was simplified to a single threshold per pad. This approach is commonly denominated as a digital readout. In the configuration with 54 active layers, the channel count was 54 × 96 × 96 = 497,664, which at the time constituted a world record in calorimetry.

## OVERVIEW OF THE ELECTRONIC READOUT SYSTEM

The electronic readout system was optimized for the readout of large number of channels and offers the option to be scaled up to the tens of millions of readout channels envisaged for a hadron calorimeter operating at a future lepton collider. A block diagram of the system is shown in Fig. 2. The electronics is divided into two parts: The "on-detector" electronics processes charge signals from the detector, collects data for transmission out, and acts as the interface for slow controls. The "back-end" electronics receives and processes the streams of data from the front-end electronics, and in turn passes it to the Data Acquisition (DAQ) system. It also serves as interface to the timing and trigger systems.

Because of the high channel count, a custom integrated circuit has been developed for the front-end electronics. The device, called DCAL III [14], performs all of the front-end processing, including signal amplification, discrimination/comparison against threshold, recording of the time hits, temporary storage of data, and data read out. It has a control interface that is used for configuring the chip, performing charge injection, etc. It services 64 detector channels with a choice of two programmable gain ranges with sensitivities of ~10 fC and ~100 fC, respectively. The chips used for the DHCAL were based on the third design iteration.

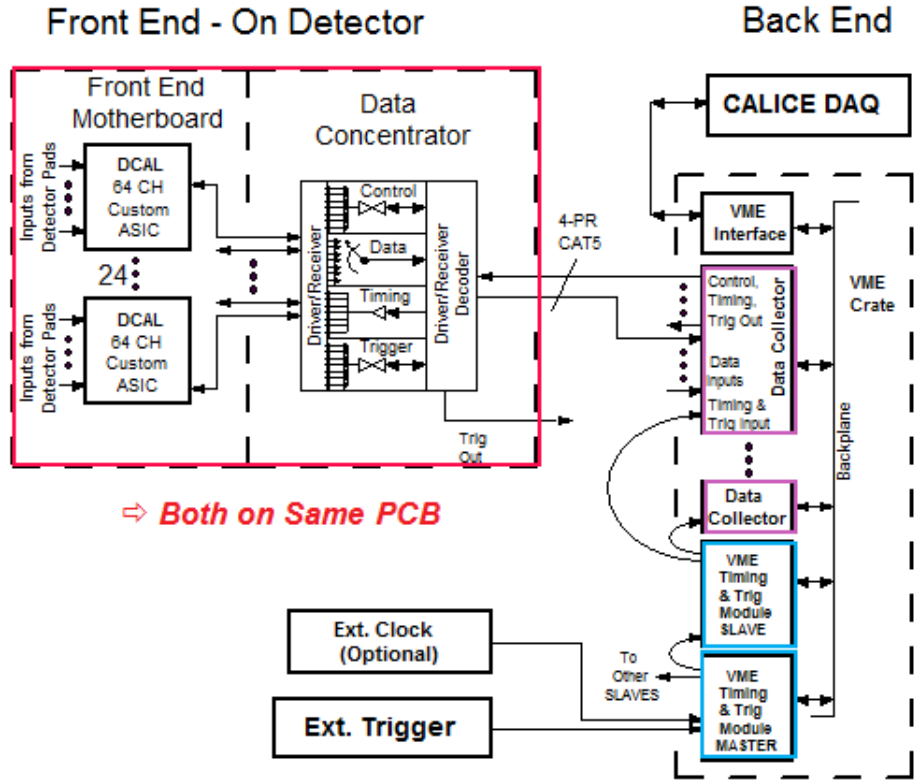

**Figure 2.** Block diagram of the electronic readout system.

In normal operation, charge is received by the front-end amplifiers of the DCAL III chip. The charge signals are amplified and shaped, and the outputs are compared to a programmable threshold (common to all channels in the chip). The resulting hit pattern is recorded along with a time. The timing of hits in the DCAL III chip is implemented using the concept of a "timestamp" counter. This counter is reset once per second across the system and advances with each 100 ns clock. The latter is synchronous across the entire system. The data are captured in a readout buffer inside the chip, either from an external trigger or self-triggered, and are read from the chip using high-speed serial bit transmission. The chip also features slow control functions, on-board charge injection, and the ability to mask off noisy channels.

The chips reside on sophisticated front-end printed circuit boards that are part of the active elements of the detector. Charge signals are received on the Pad boards, and are transmitted from the pads on the bottom side to the input pins of the DCAL chips that reside on the top side. The inner layers of the front-end board also contain the routing of the clock and digital control and data lines, as well as power and ground. There are 24 chips on each front-end board, servicing 1,536 channels. There is virtually no dead detector space due to the electronics, and the Readout boards can be tiled on three sides to cover large detector planes.

The readout operation overlaps signal acquisition, making noise performance critical. The data is read out serially from the front-end boards using "data push" into custom VME cards in the back-end system called Data Collectors. The data are time-sorted using the timestamps, and are stored in readout buffers. From there the data are read periodically into a computer, where higher-level algorithms perform the event reconstruction. In addition, the Data Collectors provide an interface to the front-ends for slow control communication and timing.

The VME crates that host the Data Collectors also contain a Timing & Trigger Module that receives timing and trigger signals from peripheral subsystems and communicates with the Data Collectors to provide this information to the front-ends. A graphical representation of the system configuration is shown in Fig. 3.

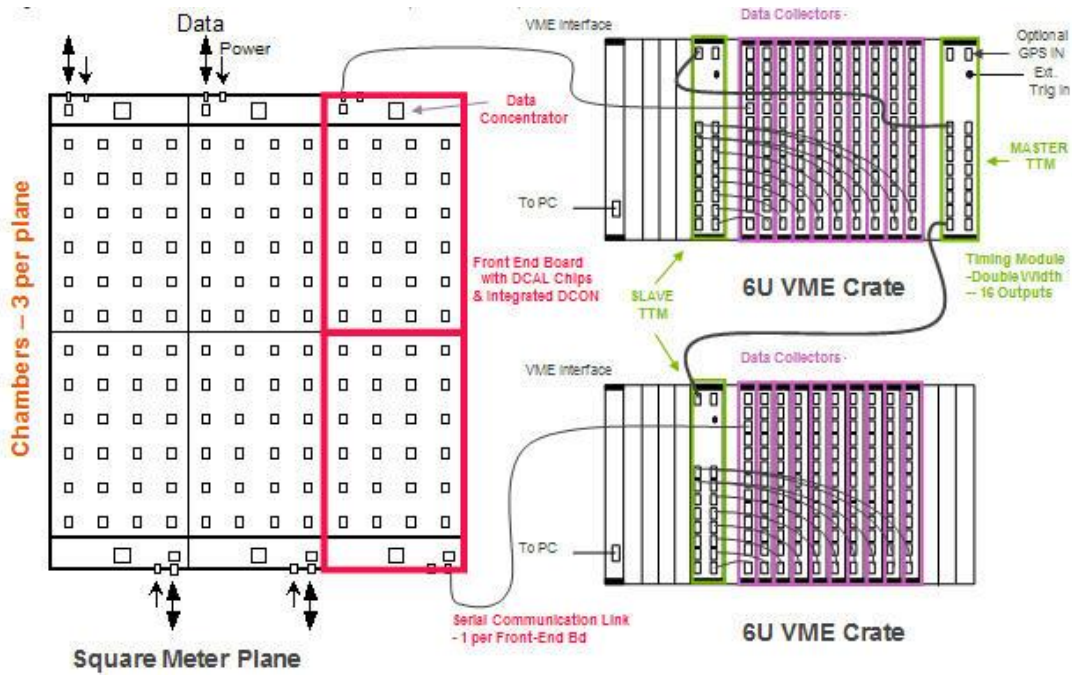

**Figure 3.** Graphical depiction of the physical connectivity of the DHCAL electronic readout system.

## DETAILED DESCRIPTION OF THE COMPONENTS OF THE READOUT SYSTEM

### a) The DCAL III chip

The DCAL III chip was designed in the TSMC 0.25 μm CMOS process. It consists of 64 channels of front-end amplification, shaping and discrimination. A block diagram of the chip is shown in Fig. 4. The following describes the chips major functions.

When a channel receives charge that exceeds a programmable threshold, the output of the amplifier/shaper/discriminator front-end produces discriminated "hits". Unwanted channels can be disabled by a Mask Register. Hit Catchers convert the discriminated pulses into digital signals exactly one clock cycle wide and delayed by one clock cycle regardless of the original magnitude or time over threshold of the input signal. A 24-bit time stamp register is associated with every time slice. With the 100 ns clock period, this means that the register can run for more than a full second before time stamp counters roll over. The Pipeline Delay can hold data for 20 time slices, or it can be bypassed depending on the setting of a configuration parameter.

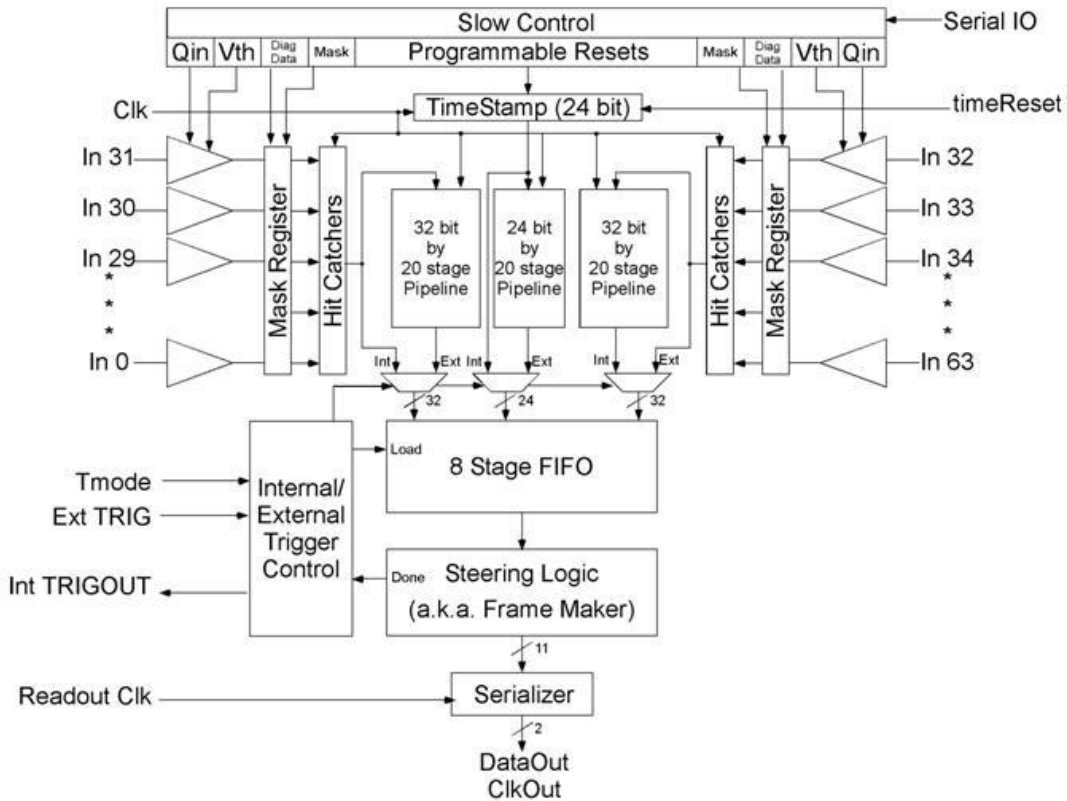

**Figure 4**. Block diagram of the DCAL III ASIC.

Data enters the 8-stage FIFO as a result of a trigger from the Trigger Control logic. The origin of that trigger can be internal or external to the chip, which again is controlled by a configuration parameter. Regardless of the origin of the trigger, once a time slice is triggered, the 88 bits of time slice data (24 bits of time stamp and 64 bits of hit/no hit information) are processed using a data push readout architecture. Whenever there is data present in the 8-stage FIFO, the FIFO and the Steering Logic work together to output that data through the Serializer. No external request is necessary. When no data is present in the FIFO, the Steering Logic and the Serializer output synchronization and status information.

Event data is written from the chip using a serial bit stream and LVDS drivers. The primary 10 MHz clock is used as the clock for the data output, transferring one bit every 100 ns. It takes 12.1 µs to transfer one time slice of data from a given chip, corresponding to a maximum event rate of 82 kHz. Each chip consumes 100 mA at 2.5 V or 3.9 mW per channel.

The chip has a slow control interface similar to an SPI interface, as seen at the top of Fig. 4. The slow controller is capable of configuring all analog and digital values needed by the chip. This includes the front-end biases, the discriminator thresholds (Vth), the calibration pulse height (Qin), and all digital control signals. The slow control also accesses the Kill and Inject registers through which a user can mask off unwanted channels (kill) and/or connect one or more channels to an on-chip pulser circuit (inject) for calibration and testing. The actual pulsing of the pulser circuit is controlled via an external digital signal that can be connected to multiple chips, resulting in a highly reproducible system calibration test.

The fabrication run of DCAL III chips was done through MOSIS [15], and was comprised of 11 wafers. The chips were packaged in a 176-pin, 24 mm × 24 mm LQFP package. A small number of chips were mounted on a test board and underwent extensive testing on the bench. The tests included all relevant aspects of the chip's operation in the DHCAL, such as slow control functions, internal and external charge injection, etc. As an example of these tests, Fig. 5 shows the response for the 64 channels of a chip to internally injected charge as function of threshold (S-curves). Notice the sharp transition from 0 to 100 counts and the tight distribution of the curves. The bench tests failed to identify a single design flaw.

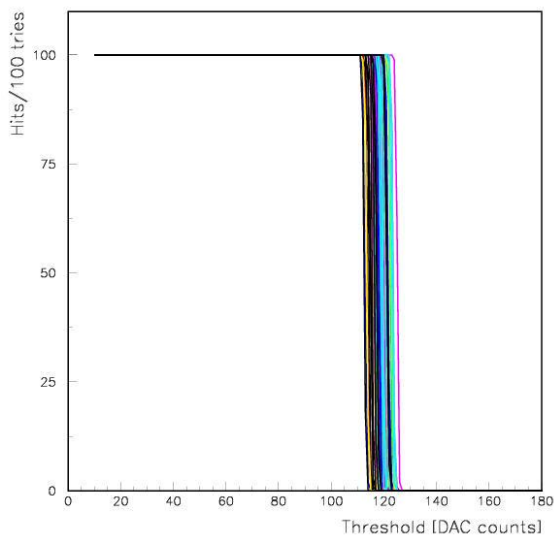

**Figure 5.** S-curves as measured with the DCAL III chips. The chip was operated in high gain mode and the charge was injected internally.

The production testing was performed at Fermilab using the ASIC Test System and Robotic Chip Tester, as shown in Fig. 6. The testing philosophy was to test every input and output on the chip and measure as many parameters as possible to assure that chips declared good, were indeed fully functional. For the DCAL III chip, all 64 input channels were exercised in 4 banks with each bank firing every 4th input. In this way, any channels which might be shorted together could have been detected. Each channel was tested using the external input and the internal charge pulser, in both high and low gain modes. Each channel's mask bit was tested, as well as the external and internal trigger modes. The current draw was measured on both the analog and digital supplies and the voltages on all of the biases and the LVDS levels of the outputs. All read/write registers were thoroughly tested and each chip was tested for all possible chip IDs and to verify that the reset loads the registers with their correct default values. This methodology assured that the chips were working as designed and ready for installation.

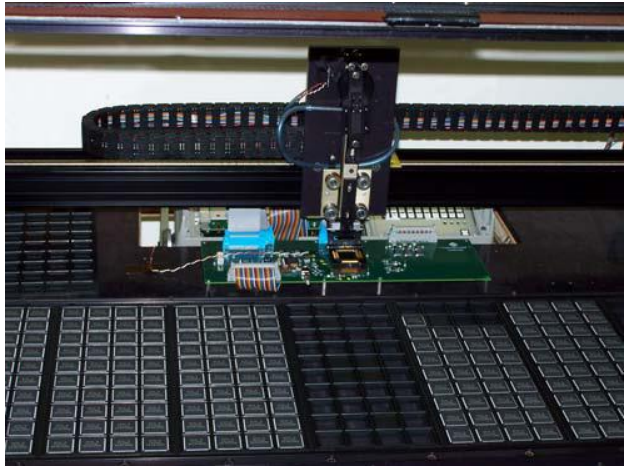

**Figure 6**. Picture of the Fermilab Robotic Chip Tester

Statistics on the type of failures were kept, and the yield for each wafer was determined. Most wafers were found to have a yield in the mid 80% range. However, one wafer had a yield of less than 50%, with most failures related to external inputs. A likely reason for the failures might be found in faulty wire bonding. The yield from the other wafers was sufficient for the required number of working parts, so the chips from the low-yield wafer were kept out of production. In the end, 10,300 chips were produced, with 8644 good parts, or 84% yield.

As the need for additional parts to instrument the tail catcher arose, the "bad" parts were retested with slightly reduced requirements to get to the required quantity of chips.

b) **Front-end boards**

The DCAL III chips reside on a Front-end board that is an integral part of the calorimeter's active elements. Each Front-end board contains 24 DCAL III chips, arranged in a $4 \times 6$ array as shown in Fig. 7. The output data streams from the DCAL chips are point-to-point serial LVDS, sending data to the Data Concentrator Field Programmable Gate Array (FPGA) seen on the left side of Fig. 7. The data stream is received by a FIFO in the Data Concentrator. A state machine in the FPGA cycles round robin through the 24 FIFOs and selects the data that has the lowest timestamp. This data is written out first, followed by the next smallest timestamp, etc. In this way, the data is time-ordered coming out of the Data Concentrator. Of course, when using an external trigger, all chips respond at the same time with the same timestamp.

The Data Concentrator also serves as interface for sending slow control information to the DCAL chips using a serial bus protocol. Each chip has a unique address on the board, allowing them to be configured uniquely. (In practice, all chips were programmed the same way.) The Data Concentrator also fans out the clock and trigger signals that are received from the Data Collector.

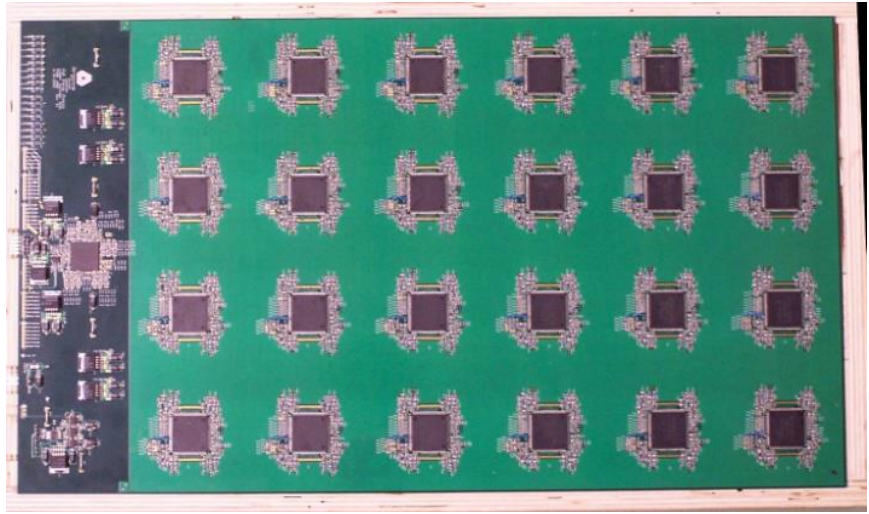

**Figure 7.** Picture of a Front-end board with 24 DCAL III chips. The darker area on the left corresponds to the Data Concentrator.

From this description, it can be seen that there is a fair amount of digital communication between the Data Concentrator FIFO on the board and the DCAL chips. This communication occurs constantly, since the serial communication lines operate continuously in order to maintain synchronization. These communication buses and control lines are designed into the Front-end board, as shown in Fig. 8. This illustrates one of the primary challenges in the design of this board. As described earlier, the charge inputs come up from the bottom side of the board to reach the chip inputs. The digital signals must be routed through this array of vias with the sensitive charge input spaced 1 cm apart in both X and Y. Noise immunity and suppression were significant challenges in this design, where the smallest signal of interest is of order 10 fC. The board was designed where all digital signals are differential LVDS, with careful matching of signal lines, strategic placement of ground planes, and careful control of common–mode return currents. Since the charge signal currents must return to the detector, this return current path must also be provided for in the design of the front-end board.

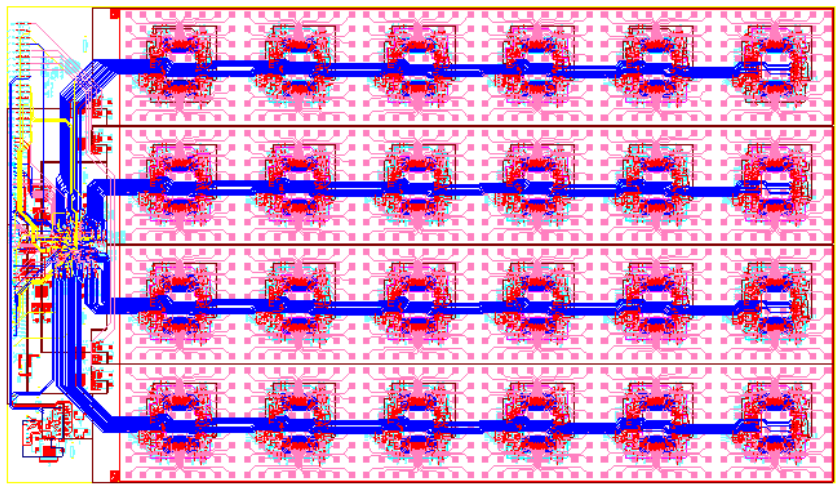

**Figure 8.** Screen shot of the Front-end board showing the internal trace layers (in blue).

c) **Gluing of the Front-end boards and Pad-boards**

The Read-out board is a sandwich, comprised of the Front-end board and the Pad board and bonded together using conductive epoxy. The reason for this is that the Pad board, which is an integral part of the detector, must not have any via holes, and must have "perfect" pads. Since there are a significant number of connections between the DCAL III chips and the internal data busses and digital signals, vias cannot be avoided in the Front-end board. Thus, there are only two design choices: either there be two boards bonded together, or there be one board that used blind and buried vias. The latter is difficult to implement on large boards due to yield problems, as well as being very expensive. Instead the simpler approach of using two boards was adopted.

The Front-end board contains 8 layers using conventional fabrication methods and materials. The Pad board is a 2-layer board, with a single via connection between each of the 1 cm$^2$ pads on the bottom and the 0.25 cm$^2$ glue pads on the top. The via holes are filled with silver epoxy to provide a smooth surface on the pad side.

A robotic gluing machine dispensed glue[1] dots onto the glue pads of the Pad board. After dispensing, the fully assembled and tested Front-end board was placed onto the Pad board and weighted down using metal blocks. To shorten the curing time the matted Front-end and Pad boards were placed overnight in an oven at $60^0$ C. The success rate of the gluing process was in excess of 99%.

d) **The Timing & Trigger Modules (TTMs)**

The Timing & Trigger Module provides support functions for the DCAL readout system. The front-end electronics needs a small number of timing signals to perform the data acquisition, and these must be synchronized over the entire detector. The system also has the provision to have the readout triggered by an external trigger system. A picture of a TTM is shown in Fig. 9.

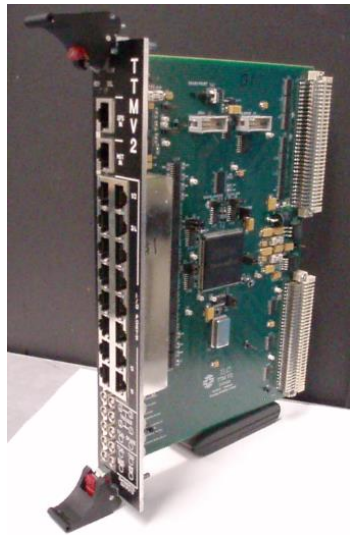

**Figure 9.** Photograph of a Timing and Trigger Module.

---

[1] Epo-tex E4011 supplied by EPOXY TECHNOLOGY INC, Billerica, MA 01821

Generally, there is one TTM in every VME crate in the system. The TTM receives timing signals from a timing source, and trigger signals from an external trigger system, and fans them out to the Data Collectors that reside in the VME crate using point-to-point connections. (The Data Collectors are described in the next section).

The system was designed to be scalable. To accomplish this, the TTM can be configured as either a Master or a Slave. When configured as a Master, the 16 outputs serve as a fan-out of signals to multiple back-end crates, in which Slave modules reside. The Master TTM accepts all external stimuli such as trigger, reset, TCAL (used for charge injection in the front-end chips,) and GPS signals. It passes these timing signals to the Slave modules which will then send out timing and control signals to the Data Collectors, which in turn send the timing signals on to the front-end electronics.

The entire DHCAL trigger and data acquisition is synchronized to a master clock with a frequency of 10 MHz. Trigger distribution and event timestamps are processed with 100 ns granularity using this master clock. The timing synchronization utilizes a pulse width-modulated serial bit stream.

The TTM printed circuit board is based on a 6U x 220 VME format and contains 8 layers with ground and power planes shielding the time and noise critical circuitry. The circuitry was designed to minimize delays and timing skew across the 16 channels that are fanned out to the DCAL system. The parts selection includes Low Skew LVDS drivers, Programmable Phase Lock Loop (PLL) and an FPGA. The FPGA allows for 100 ns communication with the master system through the reading and writing of registers. The registers contain information about the setup of the module and also timing information.

e) **The Data collectors (DCOLs)**

The Data Collector is a 6U VME module which provides several system functions: it controls a bidirectional slow control interface to the front-end electronics; it receives the event data from the front-end electronics; it interfaces to the data acquisition system through VME transactions; and it provides the means for distributing the trigger and fast clock from the timing and trigger systems to the front-end electronics. The DCOL has 13 RJ45 connectors on the front panel. One connector accepts timing signals from the TTM, while the other 12 connectors provide links to the detector front-end electronics. The module complies with the VME64 standards [16]. A picture of the module is shown in Fig. 10.

Each DCOL services 12 Readout boards using bidirectional serial communication links over 12 CAT5 cables. Two pairs in each link are used for bidirectional communications. Each pair carries an LVDS serial bit stream with a basic clock rate of 40MHz. In addition, a dedicated asynchronous isolated signal is provided to the front-end for test pulse synchronization. The links incorporate both DC and AC isolation up to several hundred MHz by means of a capacitive digital isolator, with power delivered from the DCOL via an additional cable pair.

The Front-end boards send a stream of identically formatted 16-byte packets, each containing data for one DCAL chip. The packet has a short header with Readout board and DCAL addresses, followed by the timestamp for that time slice of 100 ns data, and then the 64 bits of hit data. To send a single time slice of data from a single DCAL chip thus takes 6.4 µs. If all 24 chips on a Readout board are triggered, and all have data, the maximum event rate in reading data from the front-ends would be 6.5 kHz. In practice, not all DCAL chips on a given Front-end board contain event data, and since the Data Concentrator FPGAs on the front-ends perform zero suppression, the effective sustainable event rate is generally much higher.

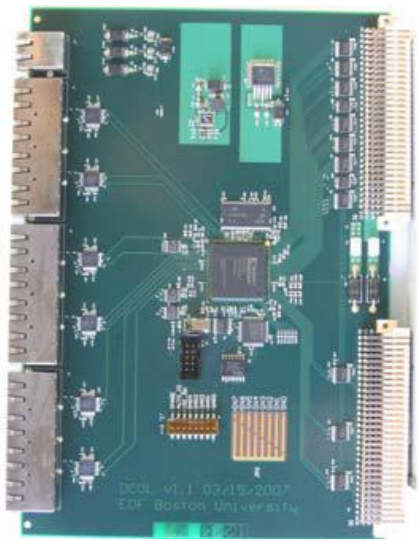

**Figure 10.** Photograph of a Data Collector.

The data received by the DCOL is stored in timestamp order in a 24Mb circular buffer. The data may be transferred over the VMEbus in various modes, including the 64-bit multiplexed block transfer mode at up to 80 Mbytes/sec. Control functions for the front-end are provided by a dedicated control channel multiplexed on the front-end links. The control functions include setting of thresholds, masking bad channels, controlling test charge injection and setting the trigger mode.

f)  **The DAQ system**

The DAQ hardware for the DHCAL consisted of two CAEN V2718 VME controllers and an A2818 PCI adapter mounted in a SuperMicro 5035B-T Workstation running Scientific Linux 5. The two VME controllers, connected in a daisy chain configuration, were read out by a 20 m optical fiber connected to the PCI adapter. The driver software and C language interface library supplied by the hardware vendor were used along with the Hardware Access Library (HAL) software library [17].

The DHCAL DAQ software was based on a modular system [18] written in C++ and originally developed to support the CALICE Readout Card (CRC) [19]. It was extended to support the DHCAL electronics. The system was built around a finite state machine. Messages, referred to as records, were passed from module to module at state transitions. The records were divided into sub-records, which may be created by modules and may add to the record to pass information along to other modules. For example, a configuration module created a sub-record containing parameters which were then extracted by a readout module that downloads the values to hardware registers. Readout modules created sub-records to hold event data read from hardware buffers and a writer module eventually wrote the entire record to a disk file. Modules are implemented in a single program or multiple programs which could run on a distributed network, with each node performing different tasks.

This system has been used by several other detector prototype projects [20,21] in the CALICE collaboration. Software modules tailored to the specific needs of the DHCAL front-end and back-end electronics were deployed in both DHCAL standalone operations, such as cosmic ray tests, as well as in beam tests along with other detectors using the CRC. In most of the beam tests a network of three computers was used: one for the DHCAL readout, one for the trigger and CRC readout, and one to record the data.

The modules developed for the DHCAL fall into one of three categories: configuration, readout and trigger. The configuration module read files containing configuration parameters and created sub-records for the readout modules. One readout module configured the registers in the front-end DCALIII chips, the Data Concentrators and the Data Collectors themselves, as well as performed the readout of the data buffers. Another readout module configured and read out the TTMs.

Different combinations of triggering and readout were used depending on the type of study. For noise studies, where the detector was self-triggered by signals above threshold, all the data collectors were read out when a timeout at a fixed interval of 1 - 10 ms had occurred. In beam tests, where the particles were delivered in spills, the externally delivered triggers were polled and counted. However, the DCOLs were only read out between spills and the event boundaries were determined from the trigger timestamps embedded in the data.

g) **System production**

A summary of the production quantities of the various system components is shown in Table I. Note that these quantities include spares and additional test stands. Generally, the system components were fabricated, assembled, and tested at various locations and delivered to Argonne for system commissioning.

**Table I**. Summary of electronic system subcomponents produced.

| Item | Number produced |
|---|---|
| DCAL III chips | 8644 |
| Readout boards | 324 |
| DCOLs | 44 |
| TTMs | 8 |
| VME crates | 5 |
| VME processors | 5 |

# PERFORMANCE OF THE READOUT ELECTRONICS

a) **Calibration of the thresholds**

The DCAL chip contains an on-board charge injection circuit. Each amplifier has its own charge injection capacitor, but there is one DAC reference voltage for all channels. By setting the DAC reference voltage and then sweeping the threshold DAC from high value to low value and looking for hits in each channel, the familiar threshold "S-curves" can be obtained, as shown in Fig. 5. The width of the plot is an indication of the uniformity, although it includes both variations in channel offsets and also variations in the individual charge injection capacitors. Since the slopes are fairly steep in the transition region, we define the threshold as the equivalent input charge that corresponds to firing the discriminator 50% of the time. By sweeping the DAC reference voltage through the entire range, the threshold gain of each amplifier can be obtained. Since this procedure uses on-chip components, the method does not provide an absolute calibration, but can be used for stability monitoring.

To obtain an absolute calibration, a precision external charge injector must be used, either at the bench level or as part of the chip check-out procedure before mounting onto the front-end boards. Measurements for one chip for both high gain and low gain are shown in Fig. 11. The nonlinearity at the high end of the range is expected, and generally the chip is not operated there. Using the first three points of the two curves, the threshold gain can be calculated. The distribution of gains for all channels in a single chip is shown in Fig. 12 and Fig. 13. From this, the uniformity of the individual channel performance can be seen, since the same charge injector was used for all measurements. The RMS of the distributions is 1.6 (3.4) % in low (high) gain mode. The average ratio comparing the low gain response to the high gain response is 5.74.

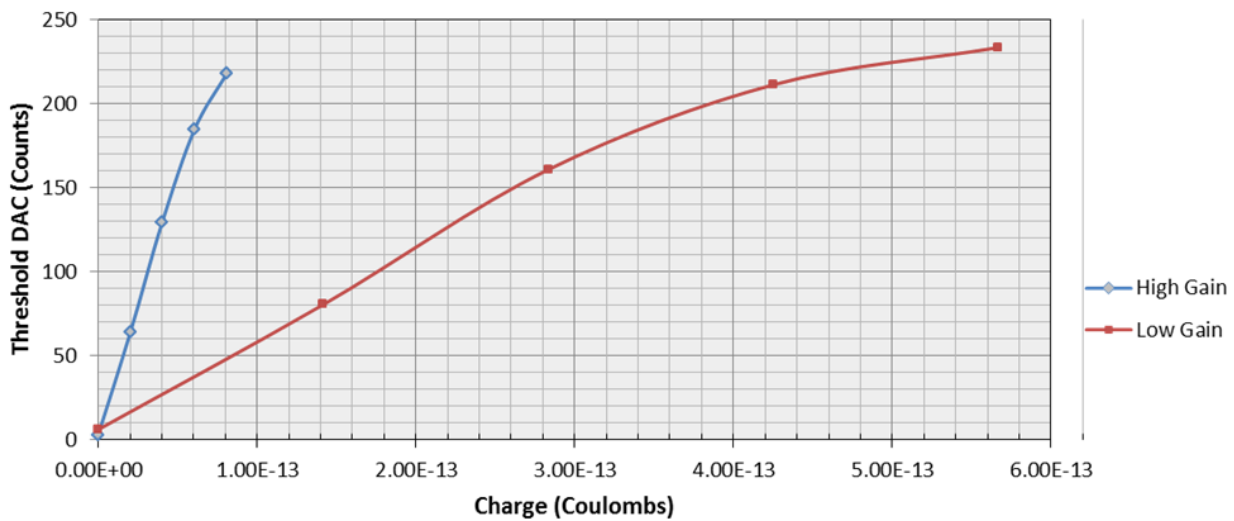

**Figure 11.** Plot of the average of the low gain and high gain responses for all channels in one chip.

a) **Noise**

The DCAL front-end circuit is nearly identical to that used in the FSSR2 chip, described in [22]. A block diagram of the front-end circuit is shown in Fig. 14. Like the DCAL amplifier, the FSSR amplifier also had two gains, 100 mV/fC and 150 mV/fC. It achieved an Equivalent Noise Charge (ENC) of 1000 electrons with a source capacitance of 20 pF. The primary change to the design for DCAL is a change of the high range (low gain) setting to 25 mV/fC to accommodate the larger charge signals produced by the RPCs.

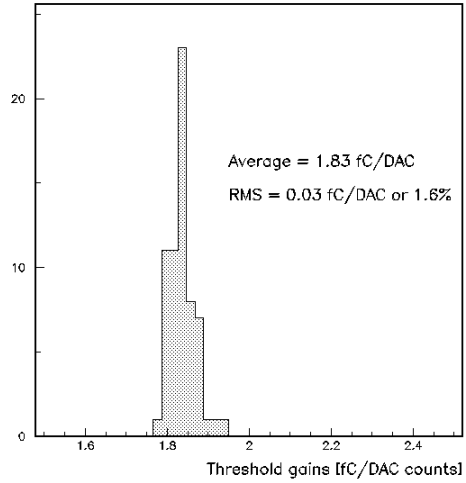

**Figure 12.** Distribution of threshold gains for the low gains of one DCALIII chip.

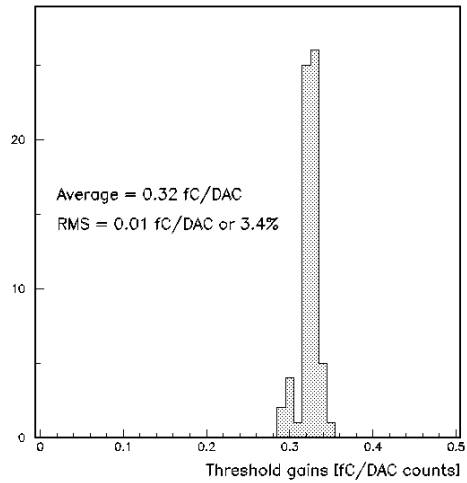

**Figure 13.** Distribution of threshold gains for the high gains of one DCALIII chip.

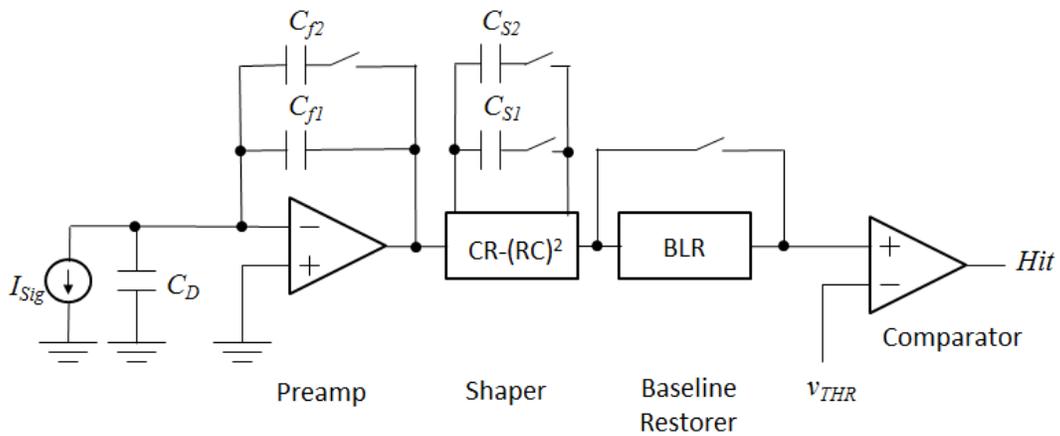

**Figure 14.** Configuration of DCAL chip front-end circuit.

Bench measurements of the noise floor of a DCAL chip are shown in Figs. 15 and 16. The measurements were made in a test stand using a test board with no source capacitance, other than the stray capacitance of the board, which was estimated to be about 10 pF. For these test, one channel was enabled at a time. The threshold DAC was initially set to full scale, and then ramped lower, one count at a time. The measurement at each DAC value consisted of 100 readings of 7 consecutive triggers, checking for hits. As before, the effective threshold is defined as the point where the response counts hits 50% of the time. The plots show the distribution of threshold values, for both low gain and high gain settings. Note that in this measurement, the noise floor as we have defined it includes not only the noise of the front-end amplifier circuitry, but also the offset voltage from the discriminator. While the threshold voltage is common to every channel, each channel has a separate discriminator, and hence the offset voltages of the discriminator contribute to the mean and spread in values. This is evident in the plots, since the non-zero entries for the low gain channels do not correspond to the high gain plot by the gain ratio. Unfortunately, because of the architecture of the chip, it was not possible to separate out these two effects. Thus, while this measurement does not show the intrinsic noise of the front-end amplifiers, it does show that the intrinsic noise floor is very close to zero DAC counts for both the high gain and the low gain channels. As will be described next, additional noise contributions from other sources dominate the overall noise performance of the system.

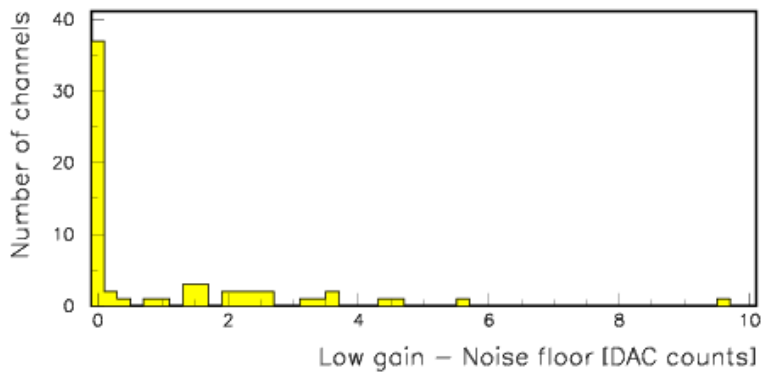

**Figure 15.** Distribution of noise floor threshold values for the low gains of one DCALIII chip.

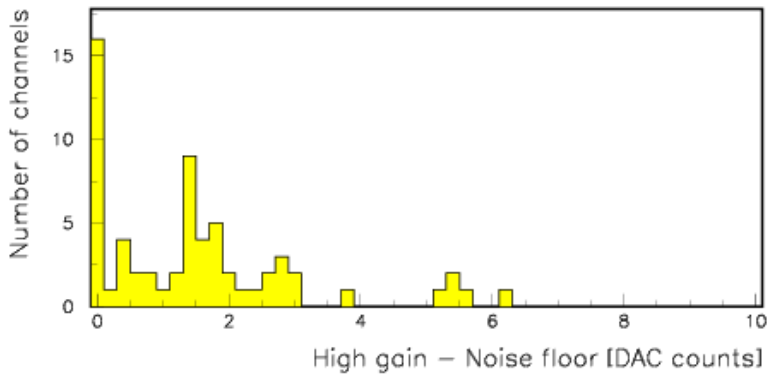

**Figure 16.** Distribution of noise floor threshold values for the high gains of one DCALIII chip.

When the chips are mounted onto the front-end boards, additional contributions are seen in the measurement of the noise floor. In the production checkout of the boards, the same noise floor measurements were performed as described above, one chip at a time. The results for one entire board are shown in Fig. 17 and Fig. 18 (24 chips, 64 channels per chip). Two effects contribute to the increase seen in the noise of the front-end board: the increase in source capacitance that is introduced by the routing of signal traces in the front-end board; and pickup from the digital traces that run through the board. Measurements of the noise on test chips as a function of source capacitance are shown in Fig. 19. It is estimated that the front-end board contributes ~100 pF of source capacitance, which results in a small contribution to the overall measured noise.

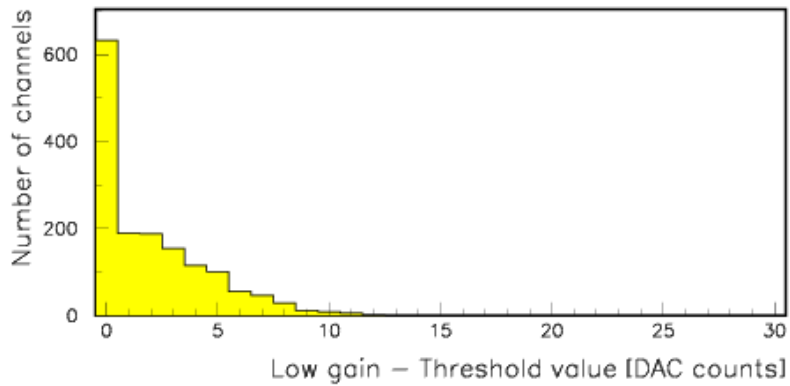

**Figure 17.** Distribution of noise floor threshold values for the low gains of one front-end board (1536 channels).

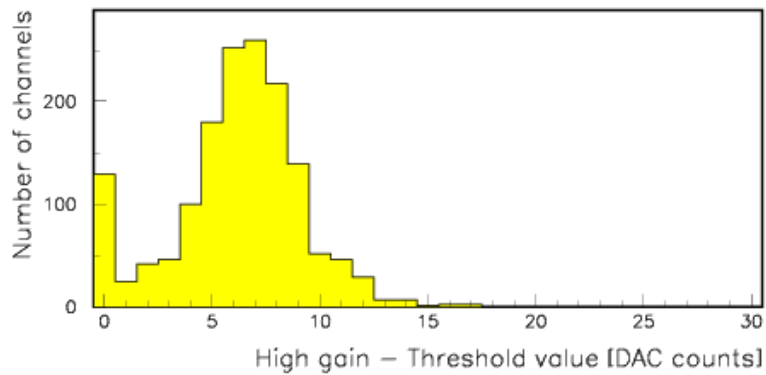

**Figure 18.** Distribution of noise floor threshold values for the high gains of one front-end board (1536 channels).

When the front-end boards are mounted onto RPC chambers and operated in the experiment, additional noise is observed. Significant factors here include additional source capacitance from the pad boards and capacitance to the ground plane of the chamber, and noise pickup from the environment. The low-gain setting was used for the RPCs, and the threshold DAC was typically set to 110 counts, which corresponds to 180 fC. This value was chosen early in the measurement program, based on lowering the thresholds as much as possible while achieving a trigger rate compatible with the bandwidth of the data acquisition system, when operated in triggerless mode. The actual noise floor, however, was significantly lower, at around 12 DAC counts, which corresponds to 22 fC in low gain. Note that this is not an RMS value. It is the value at which the discriminator fires ~50% of the time.

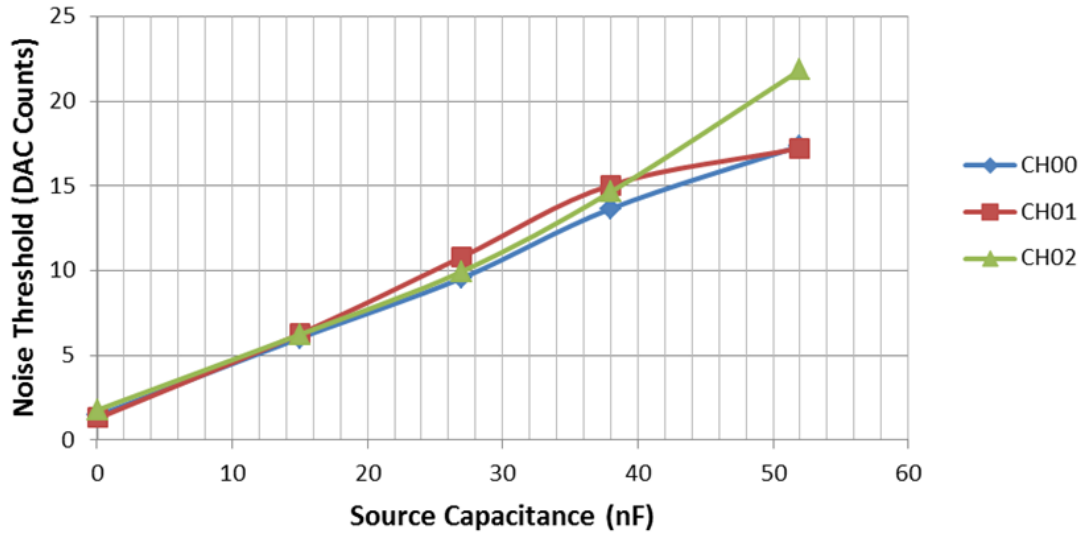

**Figure 19.** Measurement of noise versus source capacitance for three channels in a DCAL chip using high gain.

# CONCLUSIONS

The world's first large scale Digital Hadron Calorimeter (DHCAL) prototype was constructed and assembled in the period from fall 2008 to January 2011. The DHCAL utilizes Resistive Plate Chambers (RPCs) as active elements. The readout is segmented into $1 \times 1$ cm$^2$ pads, read out individually with a 1-bit resolution (=1 threshold = digital readout). Due to its fine segmentation and large size, at the time the prototype held the world record in channel counts (497,664) both for calorimetry and for RPC systems.

The prototype was extensively tested with cosmic rays and in the Fermilab test beam. It performed according to expectations. In particular, the electronic readout system, based on the DCAL III front-end chip performed exceedingly well.